\documentclass[1p]{elsarticle}
\usepackage{amssymb,amsfonts,amsmath}
\usepackage{hyperref}
\usepackage{lineno,hyperref}
\journal{Spatial Statistics}

\biboptions{authoryear}
\nolinenumbers

\begin{document}

\begin{frontmatter}

\title{Likelihood analysis for a class of spatial geostatistical compositional models}

\author{Ana Beatriz Tozzo Martins}
\address{Departament of Statistics, Maring\'a State University, Maring\'a, PR, Brazil.}
\author{Wagner Hugo Bonat\corref{mycorrespondingauthor}}
\address{Departament of Statistics, Paran\'a Federal University, Curitiba, PR, Brazil.}
\corref{mycorrespondingauthor}
\cortext[mycorrespondingauthor]{Corresponding author}
\ead{wbonat@ufpr.br}
\author{Paulo Justiniano Ribeiro Jr}
\address{Department of Statistics, Paran\'a Federal University, Curitiba, PR, Brazil.}

\begin{abstract}
We propose a model-based geostatistical approach to deal with regionalized compositions. 
We combine the additive-log-ratio transformation with multivariate geostatistical models
whose covariance matrix is adapted to take into account the correlation induced 
by the compositional structure.
Such specification allows the usage of standard likelihood methods for parameters estimation.
For spatial prediction we combined a back-transformation with the Gauss-Hermite method to approximate the conditional 
expectation of the compositions. 
We analyze particle size fractions of the top layer of a soil for agronomic 
purposes which are typically expressed as proportions of sand, clay and silt. 
Additionally a simulation study assess the small sample properties of the maximum likelihood estimator. 
\end{abstract}

\begin{keyword}
\texttt{Compositional data}, \texttt{geostatistical models}, \texttt{maximum likelihood}, \texttt{regionalized composition}.
\end{keyword}

\end{frontmatter}

\section{Introduction}
Compositional data are vectors of proportions, specifying fractions of a
whole whose elements typically sum to one or $100\%$. Given the nature 
of this data, the direct application of usual statistical techniques 
based on the Gaussian multivariate distribution on the composition 
values is not suitable. As pointed out by \cite{Aitchison:1986}, 
the constant sum constraints not only invalidate the assumption that our 
response variables are drawn from unbounded random processes, 
but also induce negative correlations between response variables.

Compositional data are frequent in earth sciences, such as, 
in mineralogy, agronomy, geochemistry and hydrology. 
In such applications, not rarely, compositions are recorded along with 
their spatial locations, and spatial patterns are of interest, 
characterizing what is called regionalized compositions \citep{Pawlowsky:1989}. 
Models accounting for spatial patterns should account for both, 
the correlation induced by the composition structure and the spatial 
correlation at a suitable scale.
 
Practical analysis of compositional data is, in general, based on the 
seminal work of \citet{Aitchison:1982} and the comprehensive monograph 
by \citet{Aitchison:1986}. \citet{Vidal:2001} presented the mathematical 
background of compositional data as equivalence classes. 
Further mathematical developments include the definitions of 
Euclidean vector space structure \citet{Pawlosky:2001} and isometric 
logratio transformation \citet{Egozcue:2003}. 
The \texttt{R} package \texttt{compositions} \citep{Boogaart:2006} 
provides a complete toolbox for analysis of compositional 
data including facilities to deal with regionalized compositions~\citep{Boogaart:2013}.

The literature about regionalized compositions is concentrated around 
the contributions of Pawlowsky 
\citep{Pawlowsky:1989, Pawlowsky:1992, Pawlowsky:1995} and its 
applications \citep{Odeh:2003, Lark:2007}. 
The monograph \citet{Pawlowsky:2004} presented the state of the art for 
the analysis of regionalized compositional data in the year $2000$.  
\citet{Tjelmeland:2003} proposed a model-based approach for the analysis 
of spatial compositional data under the Bayesian framework. 
Further developments and references can be found in \citet{Delgado:2006} 
and \citet{Pawlowsky:2015}.

The approach adopted in \citep{Pawlowsky:2004} can be summarized in 
three steps: (i) given a vector of $B$ regionalized compositions apply 
the additive-log-ratio transformation \citep{Aitchison:1986}. 
(ii) for the transformed vector use the orthodox cokriging approach 
\citep{Wackernagel:1998}. 
(iii) adopt an unbiased back-transformation to predict the compositions 
back on the original compositional scale. 
Examples of this approach with emphasis on step (iii) can be found in 
\citet{Lark:2007}. 
Such approach include the usage of traditional geostatistical techniques with 
parameter estimation based on the variogram and cross-variogram methods. 
Alternatively, a  model-based geostatistical approach 
\citep{Diggle:1998} can be considered, allowing the adoption of 
likelihood based or Bayesian statistical methods for estimation and 
prediction, inheriting related properties of consistency, asymptotic 
normality and efficiency. The application of maximum likelihood 
inference in geostatistics offers several advantages: 
The method may focus on the parameters of interest 
(sill, range, anisotropy angle, etc.). The uncertainty of the estimates
is easily assessed. The log-likelihood function may be 
used for model selection. The method is more efficient than others 
including the variogram and moments-based methods in mean square error 
terms. The method can also be used for optimum sample design. 
For more references discussing the advantages of maximum likelihood
in the context of geostatistical models, 
see \citet{Pardo:1998,Stein:1999,Diggle:2007}.

We adopt the model-based approach to deal with regionalized compositions. 
Following the aforementioned approach, we apply the additive-log-ratio 
transformation to obtain transformed response variables, for which we 
specify a common spatial component multivariate geostatistical model 
\citep{Diggle:2007}. 
For estimation of the model parameters we adopt the maximum likelihood 
method. For spatial prediction, we adopt the approach proposed by 
\citet{Pawlowsky:2004} combining a back-transformation and the 
Gauss-Hermite method to approximate the conditional expectation of the 
compositions. 
We also obtain simulations of the predictive distributions. 
Our approach produces predictions satisfying the required constant sum 
constraints and has interpretable parameters in the scale 
of the transformed response variables. 
We apply our model to analyse a data set about the distribution of 
mineral particles in the soil. 
We also present a simulation study to verify the small sample properties
of the maximum likelihood estimator.

Section~$2$ presents the compositional geostatistical model along with 
the estimation and spatial prediction procedures.
In Section~$3$ we apply the proposed model to analyse a real data set. 
Section~$4$ presents the simulation study. 
Finally, Section~$5$ provides some discussions and recommendations 
for future works. 
We provide the \texttt{R} code and data set in the supplementary 
material.

\section{The geostatistical compositional model}
In this section we describe the geostatistical compositional model 
as an extension of the bivariate Gaussian common component geostatistical 
model \citep{Diggle:2007}. 
Let $\boldsymbol{X}(\boldsymbol{u})$ be an $n \times B$ matrix of regionalized 
compositions at spatial locations 
$\boldsymbol{u} = (u_1, \ldots, u_n)^\top$ 
i.e., $\boldsymbol{X}_j(\boldsymbol{u}_i) > 0$ 
and $\sum_{j=1}^B\boldsymbol{X}_j(\boldsymbol{u}_i) = 1$ for 
$i = 1, \ldots,n$. 
Let $\boldsymbol{Y}(\boldsymbol{u})$ denote an $n \times (B-1)$ matrix of 
transformed regionalized compositions obtained by the application of the 
additive-log-ratio transformation on each row of $\boldsymbol{X}(\boldsymbol{u})$. 
Furthermore, 
let $\mathcal{Y}(\boldsymbol{u}) = (\boldsymbol{Y}_1(\boldsymbol{u})^\top, \ldots, \boldsymbol{Y}_{B-1}(\boldsymbol{u})^\top)^\top$ 
be the $n(B-1) \times 1$ stacked vector of transformed regionalized compositions by columns. 
The geostatistical compositional model assumes that $\mathcal{Y}(\boldsymbol{u})$ is 
multivariate Gaussian distributed with vector of mean 
$\boldsymbol{\mu} = (\boldsymbol{D}_1 \boldsymbol{\beta}_1^\top, \ldots, \boldsymbol{D}_{B-1} \boldsymbol{\beta}_{B-1}^\top)^\top$ and covariance matrix $\boldsymbol{\Sigma}$ given by the components,

\begin{equation}
\label{spatial}
\mathrm{Cov}(\boldsymbol{Y}_r(\boldsymbol{u}_i);\boldsymbol{Y}_{r}(\boldsymbol{u}_i)) = \sigma^2_{r} + \tau^2_{r}, \quad \quad 
\mathrm{Cov}(\boldsymbol{Y}_r(\boldsymbol{u}_i);\boldsymbol{Y}_{r}(\boldsymbol{u}_{i^\prime})) = \sigma^2_r \rho(u,\phi),
\end{equation}
and
\begin{equation}
\label{crosscov}
\mathrm{Cov}(\boldsymbol{Y}_r(\boldsymbol{u}_i);\boldsymbol{Y}_{r^\prime}(\boldsymbol{u}_{i^\prime})) = \sigma_r \sigma_{r^\prime} \boldsymbol{I_2}(i,i^\prime) + \tau_r \tau_{r^\prime} \boldsymbol{I_3}(i,i^\prime),
\end{equation}
where the functions $\boldsymbol{I}_2$ and $\boldsymbol{I}_3$ are defined by

\begin{minipage}[b]{0.48 \linewidth}
\begin{eqnarray*}\boldsymbol{I}_2(i,i^\prime)=\left\{
\begin{array}{lll}
1                                &,\,\,\text{if} & i=i^\prime,           \\
\rho(u, \phi) &,\,\,\text{if} & i\neq i^\prime,      \\
\end{array}
                           \right.
\label{eq:propostaCom}
\end{eqnarray*}
\end{minipage}\hfill
\begin{minipage}[b]{0.48 \linewidth}
\begin{eqnarray*}\boldsymbol{I}_3(i,i^\prime)=\left\{
\begin{array}{lll}
\rho_{r r^\prime} &,\,\,\text{if} & i=i^\prime,       \\
   0 &,\,\,\text{if} & i\neq i^\prime.  \\
\end{array}
\right.
\label{eq:propostaCom1}
\end{eqnarray*}
\end{minipage}
respectively. Based on this specification the $r^{th}$ component of 
the transformed regionalized compositions is given by
\begin{equation}
\label{model}
\boldsymbol{Y}_r(\boldsymbol{u}_i) = \boldsymbol{D}_r \boldsymbol{\beta}_r + \sigma_r \boldsymbol{S}(\boldsymbol{u}_i;\phi) +\tau_r \boldsymbol{Z},
\end{equation} 
where $r = 1, \ldots, B-1$. The model consists of the sum of fixed effects 
$\boldsymbol{D}_r \boldsymbol{\beta}_r$, spatially correlated 
$\boldsymbol{S}(\boldsymbol{u}_i;\phi)$ and uncorrelated 
$\tau_r \boldsymbol{Z}$ random effects. 
These effects are specified by Equation (\ref{spatial}). 
The parameters $\tau_r^2$ are sometimes called nugget effect. 
The $n \times p$ design matrix $\boldsymbol{D}_r$ contains values 
of $p$ covariates and $\boldsymbol{\beta}_r$ is a $p \times 1$ vector of 
regression parameters.

The spatial random effect $\boldsymbol{S}(\boldsymbol{u}_i;\phi)$ is a unit 
variance Gaussian random field (GRF) with correlation function 
$\rho(u;\phi)$ where $\rho \in \Re^d$ is a valid correlation function 
parametrized by $\phi$ with $d$ being the dimension of the spatial domain. 
We assume in particular correlation functions for spatially continuous process 
depending only on Euclidean distance $u = || u_i - u_{i^\prime} ||$ between 
pair of points. Popular choices are the exponential, Mat\'ern and spherical. 
At last, Equation~(\ref{crosscov}) describes the cross-covariance structure 
composed by a spatial component and a term inducing the cross-correlation, 
measured by the parameters $\rho_{rr^\prime}$. 
It is important to highlight that the range parameter is assumed common for 
all components of the transformed regionalized compositions.

\subsection{Estimation and Inference}
In this section we describe the likelihood approach used 
to estimate the model parameters.
We divide the set of parameters into two subsets, 
$\boldsymbol{\theta} = (\boldsymbol{\beta}^\top, \boldsymbol{\lambda}^\top)^\top$. 
In this notation $\boldsymbol{\beta} = (\boldsymbol{\beta}_1^\top, \ldots, \boldsymbol{\beta}_{B-1}^\top)^\top$ 
denotes a $P \times 1$ vector containing all regression parameters. 
Similarly, we let \\ $\boldsymbol{\lambda} = (\sigma_1^2 ,\ldots \sigma_{B-1}^2, \tau_1^2, \ldots, \tau_{B-1}^2,\phi,\rho_1, \ldots, \rho_{(B-1)(B-2)/2})^\top$ be a $Q \times 1$ vector of all covariance parameters. 
We use the convention to stack the correlation parameters $\rho_{rr^\prime}$ by columns. 
For a vector of observed transformed regionalized compositions $\mathcal{Y}(\boldsymbol{u})$, 
the log-likelihood function is given by,

\begin{equation}
\label{likelihood}
\mathrm{l}(\boldsymbol{\theta}; \mathcal{Y}(\boldsymbol{u})) = -\frac{(n(B-1))}{2} \ln(2\pi) - \frac{1}{2}\ln|\boldsymbol{\Sigma}| - 
\frac{1}{2}(\mathcal{Y}(\boldsymbol{u}) - \boldsymbol{D}\boldsymbol{\beta})^\top \boldsymbol{\Sigma}^{-1}(\mathcal{Y}(\boldsymbol{u}) - \boldsymbol{D}\boldsymbol{\beta}).
\end{equation}
The maximum likelihood estimator is obtained by the maximization of the log-likelihood function (\ref{likelihood}) with respect to the parameter vector $\boldsymbol{\theta}$ whose components are orthogonal \citep{Hurlimann:1992,Cox:1987}. 
For the regression parameters we can obtain a closed-form,

\begin{equation}
\label{betahat}
\boldsymbol{\hat{\beta}} = (\boldsymbol{D}^\top \boldsymbol{\Sigma}^{-1}\boldsymbol{D})^{-1}(\boldsymbol{D}^\top \boldsymbol{\Sigma}^{-1}\mathcal{Y}(\boldsymbol{u})).
\end{equation}
For the covariance parameters we adopt the \texttt{L-BFGS-B} algorithm as implemented 
in the \texttt{R} \citep{R:2015} function \texttt{optim()} for numerical maximization 
of the profile log-likelihood function obtained by substituting~(\ref{betahat}) 
in the expression~(\ref{likelihood}). 
Note that, in order to have an authorized model, i.e. positive definite covariance matrix,
we need to restrict the covariance parameters to the positive real values and the
correlation parameter in the $(-1,1)$ interval. For this reason, we used the
\texttt{L-BFGS-B} algorithm, since this algorithm allows us to introduce these 
restrictions on the parameter space. 
The algorithm requires the calculation of the score function, 
first derivative of~(\ref{likelihood}) 
with respect to the covariance parameters either numerically or analytically. 
We opt to compute the score function analytically obtaining
\begin{equation}
\frac{\partial \mathrm{l}(\boldsymbol{\theta}; \mathcal{Y}(\boldsymbol{u}))}{\partial \boldsymbol{\lambda}_q} = 
-\frac{1}{2} \mathrm{tr}\left( \boldsymbol{\Sigma}^{-1} \frac{\partial \boldsymbol{\Sigma}}{\partial \boldsymbol{\lambda}_q} \right) -
\frac{1}{2} (\mathcal{Y}(\boldsymbol{u}) - \boldsymbol{D}\boldsymbol{\hat{\beta}})^\top\left ( -\boldsymbol{\Sigma}^{-1} \frac{\partial\boldsymbol{\Sigma}}{\partial \boldsymbol{\lambda}_q} \boldsymbol{\Sigma}^{-1} \right ) (\mathcal{Y}(\boldsymbol{u}) - \boldsymbol{D}\boldsymbol{\hat{\beta}})
\end{equation}
where $\frac{\partial \boldsymbol{\Sigma}}{\partial \boldsymbol{\lambda}_q}$ 
denotes the partial derivative of $\boldsymbol{\Sigma}$ 
with respect to the element $\boldsymbol{\lambda}_q$ for $q= 1, \ldots, Q$. 
Such derivatives are easily computed using matrix calculus \citep{Wand:2002}.

Let $\boldsymbol{\hat{\theta}}$ be the maximum likelihood estimator of $\boldsymbol{\theta}$. 
Then the asymptotic distribution of $\boldsymbol{\hat{\theta}}$ is

\begin{equation}
\label{asymptotic}
\begin{pmatrix}
\boldsymbol{\hat{\beta}} \\ 
\boldsymbol{\hat{\lambda}}
\end{pmatrix} \sim 
N\left ( \begin{pmatrix}
\boldsymbol{\beta}\\ 
\boldsymbol{\lambda}
\end{pmatrix};  
\begin{pmatrix}
\mathrm{I}_{\boldsymbol{\beta}}(\boldsymbol{\hat{\beta}}) & \boldsymbol{0} \\ 
\boldsymbol{0} & \mathrm{I}_{\boldsymbol{\lambda}}(\boldsymbol{\hat{\lambda}})
\end{pmatrix}^{-1}
 \right )
\end{equation}
where $\mathrm{I}_{\boldsymbol{\beta}}(\boldsymbol{\hat{\beta}}) = \boldsymbol{D}^\top\boldsymbol{\hat{\Sigma}}\boldsymbol{D}$ and $\mathrm{I}_{\boldsymbol{\lambda}}(\boldsymbol{\hat{\lambda}})$ are the Fisher information matrices for $\boldsymbol{\beta}$ and $\boldsymbol{\lambda}$, respectively. The adopted asymptotic regime is compatible with the increasing domain framework, for further references see \citet{Chang:2013,Pardo:1998,Diggle:2007}. It is not possible to obtain a closed-form for $\mathrm{I}_{\boldsymbol{\lambda}}(\boldsymbol{\hat{\lambda}})$. Thus, we replace it by the observed information matrix obtained numerically using the Richardson method as implemented in the \texttt{R} package \texttt{numDeriv} \citep{numDeriv}.

We shall show in Section $4$ through of simulation studies that often this type of asymptotic result does not work well for covariance parameters. In the context of geostatistical analysis of compositional data we are particularly interested in the covariance parameters. 
Thus, we recommend to use the profile likelihood approach to compute confidence intervals for covariance
parameters, mainly when analysing small or medium sized data sets. 
Details about how to implement profile likelihood computations in \texttt{R} can be found in \citet{bbmle}.

\subsection{Spatial prediction}
In this section we describe the spatial prediction in the context of geostatistical compositional models. 
The procedure we shall present in this section is often called additive logistic normal kriging or 
ALN-kriging for short~\citep{Delgado:2006,Pawlowsky:2004}.
The objective is to predict the values of $\mathcal{Y}_0(\boldsymbol{u}_0)$ additional random variables at any arbitrary spatial locations $\boldsymbol{u}_{0}$ within the study region.
The best linear unbiased predictor of $\mathcal{Y}_0(\boldsymbol{u}_0)$ is the conditional expectation of $\mathcal{Y}_0(\boldsymbol{u}_0)|\mathcal{Y}(\boldsymbol{u})$ whose expression is presented in Equation~(\ref{prediction}) along with the expression for the conditional covariance. We suppress the spatial indexes for convenience.

\begin{equation}
\label{prediction}
\mathrm{E}(\mathcal{Y}_0|\mathcal{Y}) = \mathrm{E}(\mathcal{Y}_0) + 
\boldsymbol{\Sigma}_{\mathcal{Y}_0\mathcal{Y}}\boldsymbol{\Sigma}_{\mathcal{Y}\mathcal{Y}}^{-1}(\mathcal{Y} - \mathrm{E}(\mathcal{Y})), \quad \quad \mathrm{Cov}(\mathcal{Y}_0|\mathcal{Y}) = \boldsymbol{\Sigma}_{\mathcal{Y}_0\mathcal{Y}_0} - \boldsymbol{\Sigma}_{\mathcal{Y}_0\mathcal{Y}} \boldsymbol{\Sigma}^{-1}_{\mathcal{Y}\mathcal{Y}} \boldsymbol{\Sigma}_{\mathcal{Y}\mathcal{Y}_0}.
\end{equation}
In practice, the unknown parameters in the expectation and covariance structures are replaced by the maximum likelihood estimates. Note that from this procedure we obtain predictions for the stacked regionalized transformed compositions at non-observed spatial locations $\boldsymbol{u}_0$. The next objective is to back-transform these predictions to the original composition scale i.e., the unit simplex. 
For a single spatial location, let $\boldsymbol{\mu}_{\boldsymbol{Y}}$ and $\boldsymbol{\Sigma}_{\boldsymbol{Y}}$ be the expectation and covariance matrix of the additive-log-ratio transformed variable $\boldsymbol{Y}$ obtained by Equation (\ref{prediction}). The probability density function of $\boldsymbol{X}$ is given by

\begin{equation}
\label{integral}
f(\boldsymbol{X}) = (2\pi)^{-\frac{B-1}{2}} | \boldsymbol{\Sigma}_{\boldsymbol{Y}} |^{-\frac{1}{2}} 
\exp\left\{-\frac{1}{2}(\mathrm{alr}(\boldsymbol{X}) - \boldsymbol{\mu}_{\boldsymbol{Y}})^\top \boldsymbol{\Sigma}_{\boldsymbol{Y}}^{-1}(\mathrm{alr}(\boldsymbol{X}) - \boldsymbol{\mu}_{\boldsymbol{Y}}) \right\}
\left ( \prod_{i=1}^B X_i \right)^{-1},
\end{equation}
where $\mathrm{alr}(\boldsymbol{X})$ denotes the additive-log-ratio transformation applied on the vector of compositions $\boldsymbol{X}$. The Equation (\ref{integral}) is recognizable as the multivariate Gaussian distribution with an additional term, which is the Jacobian of the back-transformation \citep{Pawlowsky:2004}. By assuming that the B-part simplex is a constraint subset of the B-dimensional real space, along with the induced Euclidean geometry and Lebesgue measure, an unbiased predictor of $\boldsymbol{X}$ can be obtained by computing the expectation of $\boldsymbol{X}$ i.e.,
\begin{equation}
\label{predictions}
\mathrm{E}(\boldsymbol{X}) = \int \boldsymbol{X} f(\boldsymbol{X}) d\boldsymbol{X}, 
\end{equation}
we adopt the Gauss Hermite method to solve the intractable integral. 
Basically, the Gauss Hermite method changes the intractable integral by a weighted finite sum,

\begin{equation}
\label{Gauss}
\int_{\Re^{B-1}} f(\boldsymbol{G}) \exp \left \{ -\boldsymbol{G} \boldsymbol{G}^\top \right \} d \boldsymbol{G} \approx \sum_{i_1 = 1}^K  \ldots \sum_{i_{(B-1)}=1}^K w_{i_1}, \ldots, w_{i_{(B-1)}} f(G_{i_1}, \ldots, G_{i_{(B-1)}}),
\end{equation}
where $K$ is the number of points used for the approximation, $\boldsymbol{G}$ are roots of the Hermite polynomial $H_k (\boldsymbol{G})(i = 1 < 2, . . . , K)$ and $w_i$ are weights given by,
\begin{equation*}
w_i = \frac{2^{K-1}K!\sqrt{\pi}}{K^2[H_K(G_i)]^2}.
\end{equation*}
The Gauss Hermite method is easily implemented in \texttt{R}, as the function \texttt{gauss.quad()} from package \texttt{statmod} \citep{statmod:2013} provides the weights and the Gauss Hermite points. \citet{Pawlowsky:2004} show that the auxiliary function $f(\boldsymbol{G})$ required in Equation (\ref{Gauss}) is given by,
\begin{equation}
f(\boldsymbol{G}) = \pi^{-\frac{B-1}{2}} \mathrm{agl}(\sqrt{2} \boldsymbol{R}^\top \boldsymbol{G} + \boldsymbol{\mu}_{\boldsymbol{Y}}),
\end{equation}
where $\mathrm{agl}$ denotes the additive generalized logistic (back-transformation) and $\boldsymbol{R}$ denotes the Cholesky decomposition of $\boldsymbol{\Sigma}_{\boldsymbol{Y}}$. Note that, the $\mathrm{agl}$ back-transformation guarantees that the predicted values satisfy the required constant sum constraints~\citep{Odeh:2003}. A related matter corresponds to the calculation of confidence intervals for the predicted compositions. There is no straightforward way to build consistent and optimal confidence intervals for the predicted compositions.
In general, it is due to the assumed space structure of the data support, i.e. bounded values, modelled by assuming the additive logistic normal distribution, which in turn assumes to be embedded in the whole real space, with the usual real space structure and Lebesgue measure.
For a detailed discussion about the limitations of ALN-kriging as well as assumed conditions, we refer the interested reader to \citep{Delgado:2006}. In this paper, we do not pursue in this problem, however, we present an alternative approach based on Monte Carlo simulation of the predictive distribution.

Our approach follows the lines of~\citep{Tjelmeland:2003}, thus for estimated $\boldsymbol{\mu}_{\boldsymbol{Y}}$ and $\boldsymbol{\Sigma}_{\boldsymbol{Y}}$ simulating values from this 
multivariate Gaussian distribution is straightforward. We denote simulated values by $\boldsymbol{Y}_s$. 
We apply the back-transformation on the simulated values to obtain values $\boldsymbol{X}_s$. 
An unbiased predictor of $\boldsymbol{X}$ is the sample mean of $\boldsymbol{X}_s$.
An appealing feature is that prediction of other quantities of interest, linear and non-linear can be also obtained 
applying the functional of interest to the simulated values. Furthermore, confidence intervals 
for the predicted compositions can be replaced by intervals based on quantiles, which in turn are easily obtained based on the simulated values.

\section{Data analysis}
\begin{figure}[htbp]
\centering
\includegraphics{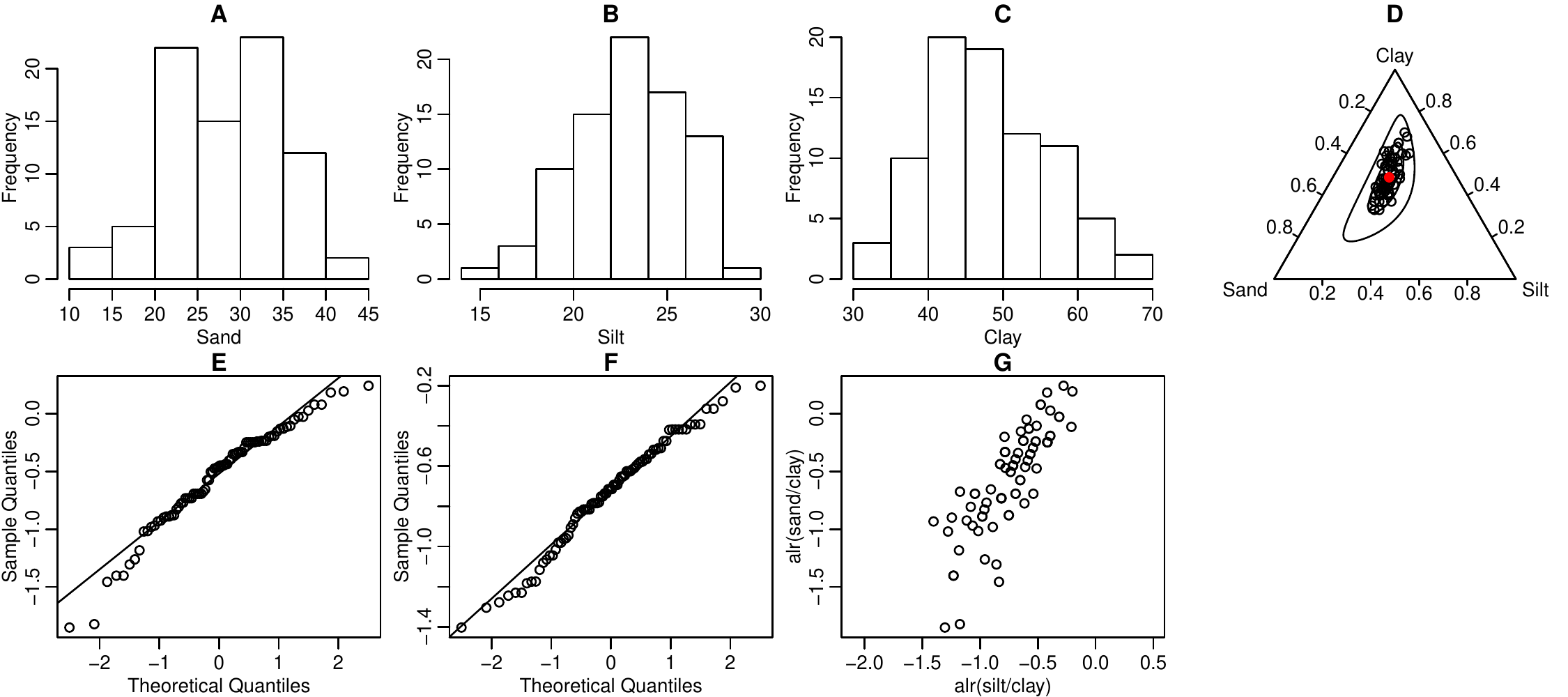}
\caption{Histograms (A to C) for the proportions of sand, silt and clay.
Ternary diagram and geometric mean (red dot) of particle size fractions (D). 
Normal quantile plot for the additive-log-ratio (alr) transform of sand (E) and silt (F) contents, 
with clay as the denominator of the ratio. 
Scatterplot relating the alr transform of sand over clay with alr transform of silt over clay (G).}
\label{fig:descritiva}
\end{figure}

In this section we report analysis of particle size fractions of sand, 
silt and clay measured at an experimental plot within the Are\~ao 
experimental farm belonging to the Escola Superior de Agricultura 
Luiz de Queiroz, Piracicaba, S\~ao Paulo State, Brazil.
The soil was sampled in the soil layer of $0$ to $20$ centimetres 
at $82$ points and on a regular grid with 20 metres spacing, 
Figure~\ref{fig:descritiva} shows the data as a ternary diagram,
histograms for each component of the composition along with
normal quantile plot and a scatterplot for the transformed response
variables.

\setkeys{Gin}{width=0.99\textwidth}

The silt fractions have smaller values and variability whereas clay is 
the predominant component with the largest variability.
In general the Normal quantile plots (E and F) support our assumption of 
marginal Gaussian distribution for the transformed response variables. 
The scatterplot shows a clear positive correlation between the transformed
response variables.
This example illustrates a fairly common situation in agriculture 
where the application of geostatistical compositional model is required. 
The distribution of mineral particles between size fractions, typically 
sand, silt and clay, affects many properties of the soil, 
such as water relations, chemistry, organic carbon dynamics and 
mechanical properties. In general the main goal of this type of spatial 
analysis is to predict the particle size fractions of the soil at a grid 
covering the area to define areas for possible different management 
practices. Interest can be in mean values, maxima and minima 
as well as exceedance of critical values. 

We computed the additive-log-ratio transform of sand and silt contents, 
with clay, the most abundant content, as the denominator of the ratio. 
For the transformed regionalized compositions we fitted the 
geostatistical compositional model with exponential correlation function. 
Parameter estimates, standard errors (SE) and asymptotic $95\%$ 
confidence intervals are shown in Table \ref{tab:resultado}.

\begin{figure}[htbp]
\centering
\includegraphics{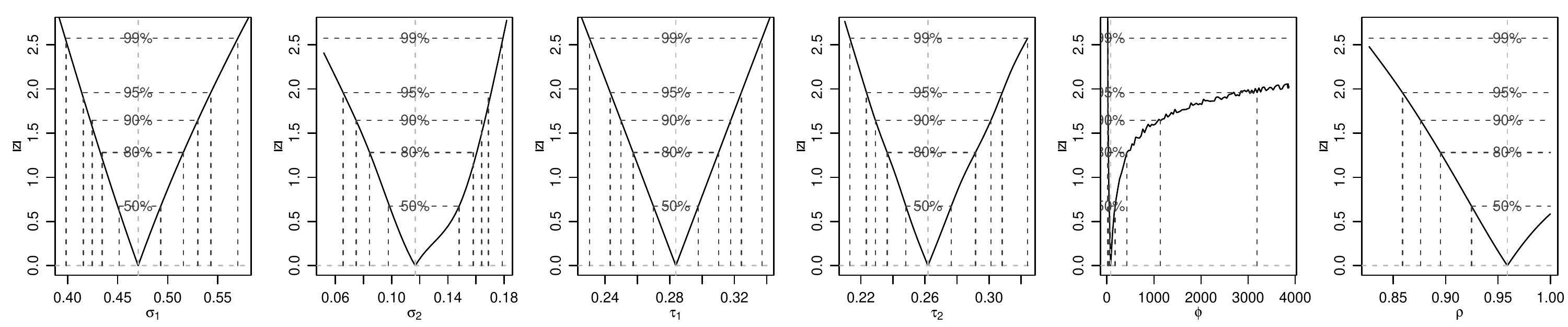}
\caption{Profile likelihoods for covariance parameters.}
\label{fig:perfil}
\end{figure}

\begin{table}[h]
\center
\caption{Parameter estimates, standard errors (SE) and asymptotic $95\%$ confidence intervals.}
\label{tab:resultado}
\begin{tabular}{crrrr} \hline 
Parameter   & Estimate  & SE      & $2.5\%$     & $97.5\%$   \\ \hline
$\beta_1$   & $-$0.7864 &0.2561  & $-$1.2883    & $-$0.2845  \\
$\beta_2$   & $-$0.7943 &0.0694  & $-$0.9304    & $-$0.6583  \\
$\sigma_1$  &0.4705     &0.1827   &0.1125       &0.8285      \\
$\sigma_2$  &0.1168     &0.0690   & $-$0.0185   &0.2520      \\
$\tau_1$    &0.2838     &0.0491   &0.1875       &0.3800       \\
$\tau_2$    &0.2619     &0.0220   &0.2187       &0.3050       \\
$\phi$      &81.4365    &80.4313  & $-$76.2059  &239.0789     \\
$\rho$      &0.9589     &0.0559   &0.8492       &1.0685     \\ \hline  
\end{tabular}
\end{table}
Following the notation introduced in the Section $2$, $\beta_1$ and $\beta_2$
denote the means for the transformed response variables~\texttt{alr(Sand/Clay)} and 
\texttt{alr(Silt/Clay)}, respectively. 
Results on Table~\ref{tab:resultado} show that the variability of the 
first transformed component is larger than of the second. 
The cross-correlation is large and the proportion of variability 
attributed to the spatial effect is larger for the first component. 
The value of the common range parameter $\hat{\phi} = 81.43$ indicates 
the presence of spatial structure, although the asymptotic confidence 
interval include artefactual negative values. 
Artefactual negative values are also included in the confidence interval 
for $\sigma_2$. It is a well-known result that, in general, the 
asymptotic result (\ref{asymptotic}) does not work well for covariance 
parameters, specially with small data sets, such are the data 
considered here. We recommend to use the profile likelihoods to quantify 
the uncertainty associated with these estimates. 
Figure~\ref{fig:perfil} shows profile likelihoods expressed in terms of 
the square root of the profile deviances for the covariance parameters 
in the geostatistical compositional model considered here.

\setkeys{Gin}{width=0.99\textwidth}

\begin{figure}[htbp]
\centering
\includegraphics{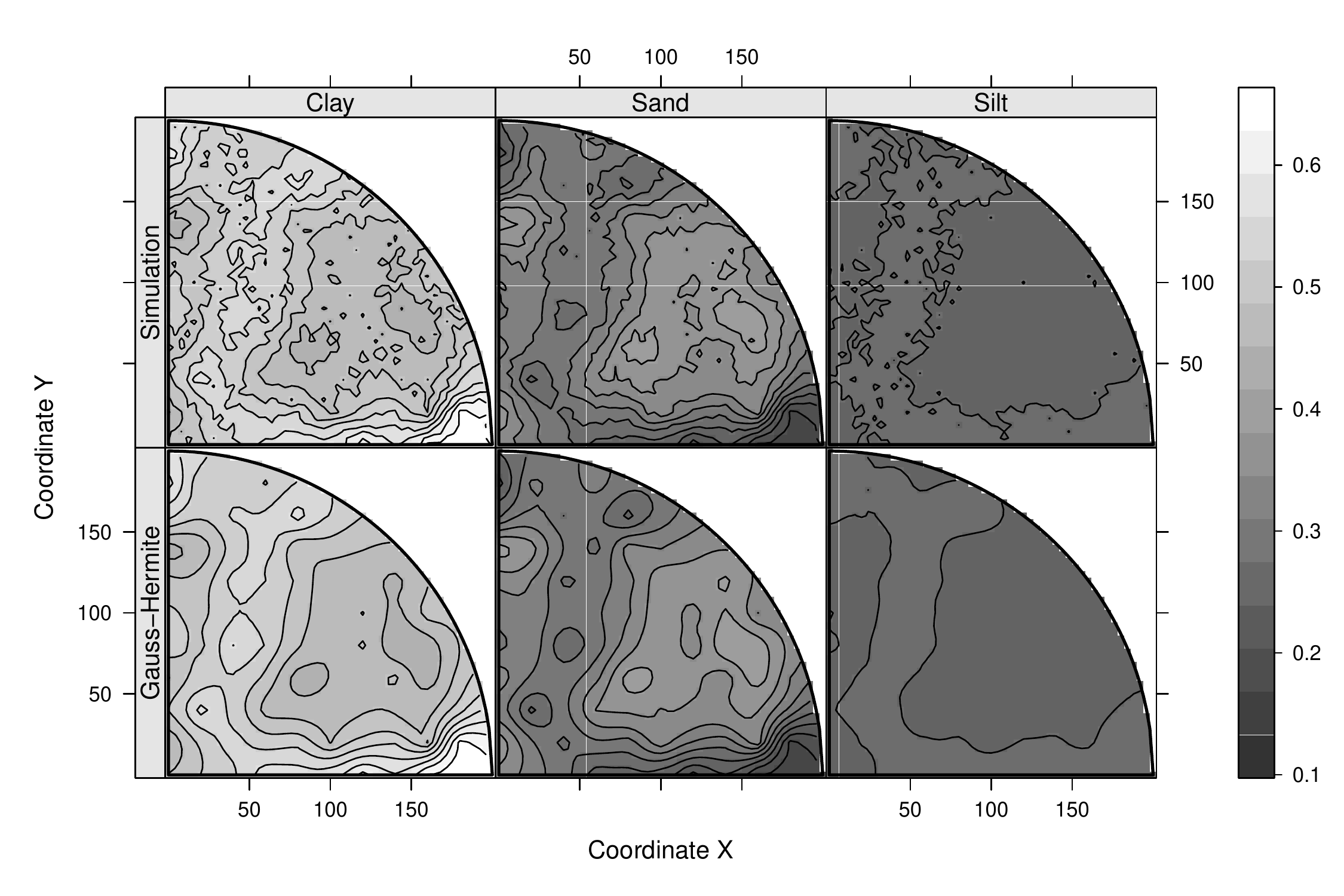}
\caption{Prediction maps of particle size fraction of clay, silt and sand by Gauss Hermite and simulation methods.}
\label{fig:prediction}
\end{figure}

The plots in Figure~\ref{fig:perfil} are compatible with a quadratic 
profile likelihood except for the range parameter $\phi$ that show a 
heavy right tail. The results confirm the worth of the spatial effect. 
Based on the fitted model and using the two methods described in 
Section $2.2$ we perform the spatial prediction of the compositions. 
Maps of predicted values are shown in Figure~\ref{fig:prediction}.

\setkeys{Gin}{width=0.99\textwidth}
In general the results returned by the two approaches agree.
Predictions based on simulations are a noisy version of the ones 
obtained with the Gauss-Hermite method.

\begin{figure}[htbp]
\centering
\includegraphics{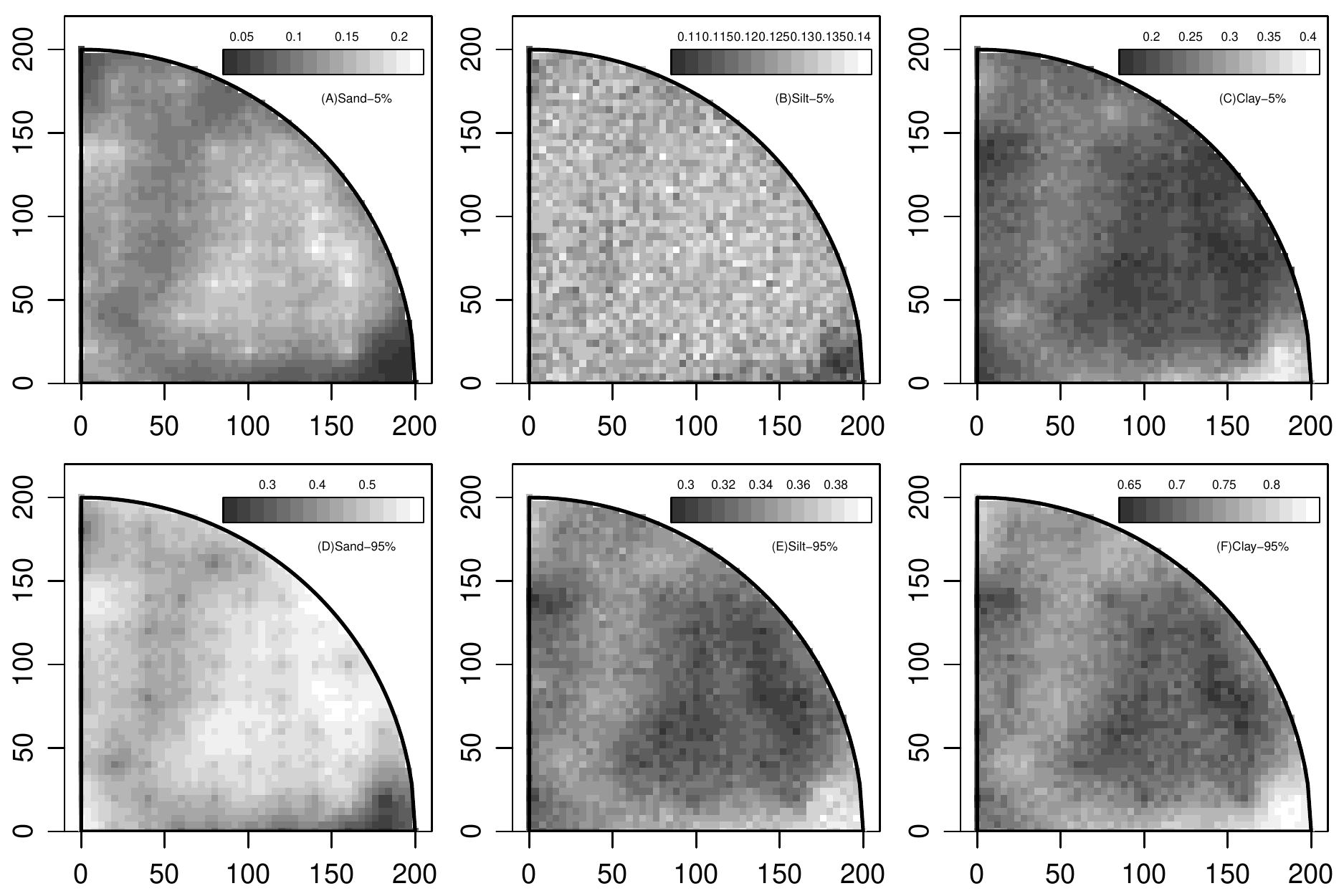}
\caption{Prediction maps of the $5\%$ and $95\%$ quantile values of particle size fraction of sand, silt and clay.}
\label{fig:maxmin}
\end{figure}

The results obtained by Monte Carlo methods are reassuring in the sense 
they validate the integral approximations.
Furthermore, they allow for computing not only predicted means and 
variances but also general predictands which otherwise would be prohibited 
by analytical methods. A typical example is the prediction of non-linear 
functions of the underlying fields. 
In order to illustrate this fact, we show in Figure~\ref{fig:maxmin} the 
prediction maps of the $5\%$ and $95\%$ quantile values for the soil fractions. 
Such quantities can be even more important than the means for defining 
soil classifications and management.

\setkeys{Gin}{width=0.99\textwidth}

\section{Simulation study}
\begin{figure}[htbp]
\centering
\includegraphics{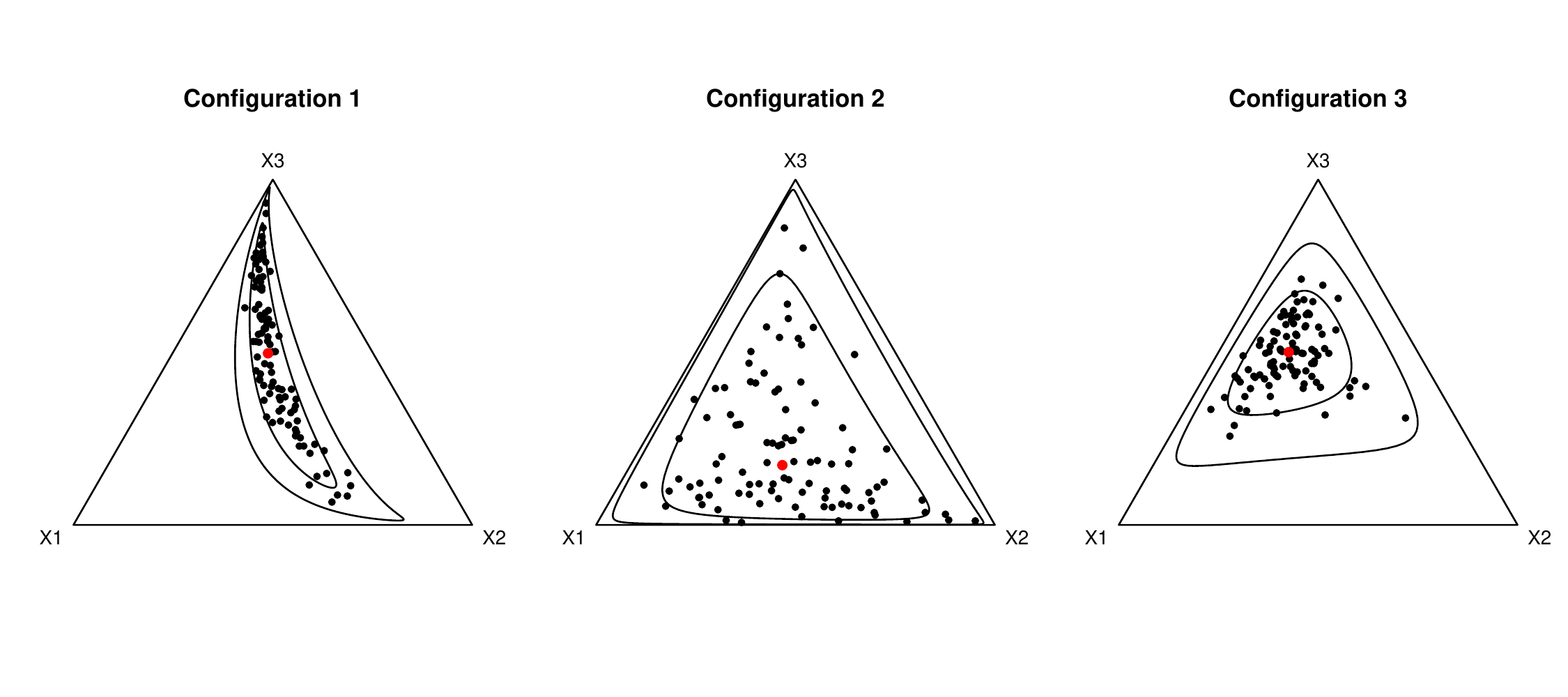}
\caption{Ternary diagrams and geometric mean (red dot) of data simulated from each parameter set 
considered in the simulation study.}
\label{fig:ternario}
\end{figure}

We now turn to a simulation study to evaluate the bias and coverage rate 
of the maximum likelihood estimators in the context of geostatistical 
compositional models. We simulated $1000$ data sets considering 
compositions with $B = 3$ components along with two sample sizes 
$n = 100$ and $n = 250$.
We show results for data simulated on a regular grid within the unit 
squared and adopt the exponential correlation function. 
We also consider three parameter configurations, in order to obtain 
different patterns of the compositional data. 
Table~\ref{tab:configuration} presents the parameter values and 
Figure~\ref{fig:ternario} shows ternary diagrams for one sample of each 
of the configurations. 

\setkeys{Gin}{width=0.7\textwidth}

\begin{table}[h]
\centering
\caption{Parameters values used in the simulation study.}
\label{tab:configuration}
\begin{tabular}{ccccccccc} \hline
Configuration         & $\beta_1$ & $\beta_2$ & $\sigma_1$ & $\sigma_2$ & $\tau_1$ & $\tau_2$ & $\phi$ & $\rho$ \\ \hline
1                     & $-$0.2  & $-$0.5  & 1        & 1.5      & 0.3    & 0.3    & 0.25 & 0.9   \\
2                     & 1       & 1       & 1.2      & 1.5      & 0.9    & 0.5    & 0.25 & 0.5   \\
3                     & $-$0.5  & $-$1    & 0.45     & 0.13     & 0.3    & 0.5    & 0.1  & 0    \\ \hline
\end{tabular}
\end{table}
The configurations $1$, $2$ and $3$ generate samples which displays the 
ternary diagram: concentrated in the middle, 
spread all over and concentrated on the left side, respectively. 

For each parameter configuration we simulated $1000$ data sets and 
fit the geostatistical compositional model proposed in Section~$2$. 
The confidence intervals were obtained using the asymptotic 
result~(\ref{asymptotic}). Table~\ref{tab:simulation} presents the bias 
and coverage rate by sample size and parameter set.

\begin{table}[h]
\caption{Bias (BS) and coverage rate (CR) by sample size and parameter set.}
\label{tab:simulation}
\begin{tabular}{c|cc|cc|cc|cc|cc|cc}
\hline 
\multicolumn{1}{l}{} & \multicolumn{4}{c}{Configuration 1} & \multicolumn{4}{c}{Configuration 2} & \multicolumn{4}{c}{Configuration 3}\tabularnewline
\cline{2-13} 
\multicolumn{1}{l}{} & \multicolumn{2}{c}{n = 100} & \multicolumn{2}{c}{n = 225} & \multicolumn{2}{c}{n = 100} & \multicolumn{2}{c}{n = 225} & \multicolumn{2}{c}{n = 100} & \multicolumn{2}{c}{n = 225}\tabularnewline
\hline
\multicolumn{1}{c}{Parameter} & BS & \multicolumn{1}{c}{CR} & BS & \multicolumn{1}{c}{CR} & BS & \multicolumn{1}{c}{CR} & BS & \multicolumn{1}{c}{LC} & BS & \multicolumn{1}{c}{CR} & BS & CR\tabularnewline
\hline
$\sigma_1$ & -.067 & .909 & -.050 & .853 & -.035 & .966 & -.059 & 0.912 & -.048 & .814 & .009 & .894\tabularnewline
$\sigma_2$ & -.101 & .901 & -.077 & .834 & -.046 & .974 & -.072 & 0.916 & -.005 & .962 & .005 & .957\tabularnewline
$\tau_1$ & .004 & .881 & -.026 & .878 & -.120 & .964 & -.051 & 0.966 & -.039 & .788 & -.062 & .945\tabularnewline
$\tau_2$ & -.004 & .891 & -.042 & .897 & -.179 & .960 & -.073 & 0.980 & -.021 & .956 & -.007 & .958\tabularnewline
$\phi$ & -.001 & .868 & -.019 & .776 & -.029 & .732 & -.037 & 0.718 & .039 & .902 & -.004 & .807\tabularnewline
$\rho$ & -.021 & .868 & .004 & .931 & -.400 & .926 & -.108 & 0.978 & -.109 & .924 & -.134 & .973\tabularnewline
\hline
\end{tabular}
\end{table}

The results show that the maximum likelihood estimators underestimate 
the covariance parameters at all cases. 
The largest bias appears in the configuration~2 and for the 
cross-correlation parameter. 
In general, as expected, the bias decreases when the sample size 
increases. For most cases the coverage rate is slightly smaller than the 
expected nominal level ($95\%$) with worse results for the range parameter.   

\section{Discussion}

We presented a model-based geostatistical approach to deal with 
regionalized compositional data. The model combines the 
additive-log-ratio transformation and multivariate geostatistical 
models whose covariance structure was adapted to take into account 
for the correlation induced by the compositional structure. 
This allows for the use of standard likelihood methods for estimation 
of the model parameters. A critical point in the analysis of 
regionalized compositional data is the spatial prediction. 
We adopted the approach proposed by~\citet{Pawlowsky:2004} 
combining a back-transformation and the Gauss-Hermite method to 
approximate the conditional expectations. 

We also obtain simulations of the predictive distribution which can 
be used for assessing quality of the results given the analytical 
approximation of the back-transformation and, possibly more important, 
to obtain predictions of general functionals of interest.
The simulation approach also provides a straightforward 
way to estimate the conditional probability of discrete textural classes
\citet{Lark2012}. Results of the predictions returned by our model 
satisfies the required constant sum constraints.

We applied the geostatistical compositional model to analyse a data set 
about particle size fractions of sand, silt and clay. 
In general, in this type of analysis the main goal is to obtain 
predictions for the fractions in a form of a map covering the study area. 
We showed through the data set that the two presented prediction methods 
provide similar and reasonable results. Through a simulation study we 
showed that in general the maximum likelihood estimators have a 
small negative bias for the covariance parameters. 
The coverage rate is slightly smaller than the expected nominal level. 
Thus, we recommend to use the profile likelihood approach to quantify 
the uncertainty associated with these estimates, 
mainly when analysing small and medium data sets.

The proposed model is based on the additive-log-ratio transformation. 
Thus, the last component in the composition is treated differently from 
the other elements. In our data set example, we opted to take the clay as 
the denominator of the ratio. Such a choice was based on the fact that the 
clay is the most abundant component of our compositions, which in turn
makes the computation of the additive-log-ratio more stable computationally.
We also fitted the models using sand and silt as the denominator of the ratio.
In general, the predictions were virtually the same. However, further studies are
required to conclude if it was a particular result for our data set or a general
feature of the model based on the additive-log-ratio transformation. 

Possible topics for further investigation and extensions include 
to improve the \texttt{R} code, and to increase the flexibility of the spatial 
modelling by allowing alternatives correlation functions, as the Mat\'ern and spherical.
Also, there is scope for the development of model-based inferential approach for 
models based on the Aitchison geometry such as presented in \citet{Pawlowsky:2015}.
The computational overhead are due to computations with the dense variance-covariance matrix. 
This overhead may be alleviated by adopting methods such as covariance 
tapering~\citep{Furrer:2006, kaufman:2008}, 
predictive processes~\citep{Eidsvik:2012}, 
low rank kriging~\citep{Cressie:2008} and SPDE models~\citep{Lindgren:2011}.

\section*{References}
\bibliography{Martins2016}

\section*{Supplementary material}

\texttt{http://leg.ufpr.br/doku.php/publications:papercompanions:compositional}

\end{document}